\documentclass[aps,superscriptaddress,twocolumn,nofootinbib,floatfix]{revtex4}%showkeys,
\usepackage{amssymb}
\usepackage{amsmath}
\usepackage{graphicx}
\usepackage{verbatim}
\usepackage[normalem]{ulem}
\usepackage[dvips]{color}
\usepackage{multirow}
\usepackage{dcolumn}                 % Align table columns on decimal point
\usepackage[bookmarksnumbered,bookmarksopen,colorlinks,citecolor=blue,linkcolor=blue]{hyperref} %
\usepackage{CJK}                     % chinese fonts
\usepackage[normalem]{ulem} % \sout{old text} for strikeout
\usepackage[dvips]{color} % For blue in-text comments and additions
\renewcommand\sout{\bgroup \color{red} \ULdepth=-.5ex \ULset}

\usepackage{epstopdf, color}

\def\es0{$E_{sym}(\rho_0)$~}
\begin{document}
\begin{CJK*}{GBK}{song}
\title{Proton-proton momentum correlation function as a probe of the high momentum tail of the nucleon momentum distribution}
\author{Gao-Feng Wei}\email[Corresponding author. E-mail: ]{wei.gaofeng@gznu.edu.cn}
\affiliation{School of Physics and Electronic Science, Guizhou Normal University, Guiyang 550025, China}
\affiliation{Guizhou Provincial Key Laboratory of Radio Astronomy and Data Processing, Guizhou Normal University, Guiyang 550025, China}
\author{Xi-Guang Cao}
\affiliation{Shanghai Advanced Research Institute, Chinese Academy of Sciences, Shanghai 201210, China}
\affiliation{Zhangjiang Laboratory, Shanghai 201210, China}
\affiliation{Shanghai Institute of Applied Physics, Chinese Academy of Sciences, Shanghai 201800, China}
\author{Qi-Jun Zhi}
\affiliation{School of Physics and Electronic Science, Guizhou Normal University, Guiyang 550025, China}
\affiliation{Guizhou Provincial Key Laboratory of Radio Astronomy and Data Processing, Guizhou Normal University, Guiyang 550025, China}
\author{Xin-Wei Cao}
\affiliation{School of Mechanical and Material Engineering, Xi'an University of Arts and Sciences, Xi'an 710065, China}
\author{Zheng-Wen Long}
\affiliation{College of Physics, Guizhou University, Guiyang 550025, China}

\begin{abstract}
Within an improved transport model, we examine effects of the high momentum tail of the nucleon momentum distribution induced by short-range correlations on the proton-proton momentum correlation function in $^{197}$Au+$^{197}$Au collisions at 400 MeV/nucleon. It is found that the proton-proton momentum correlation function from preequilibrium emissions responds sensitively to the presence as well as fraction of nucleons in the high momentum tail of the nucleon momentum distribution, but is almost robustly insensitive to other factors including the symmetry energy and the uncertainty of cutoff value of nucleon effective high momentum. In terms of the sensitivity and clearness, we propose that the proton-proton momentum correlation function from preequilibrium emissions can be as an effective probe of the high momentum tail of the nucleon momentum distribution.
\end{abstract}

%\pacs{41.20.-q, %Applied classical electromagnetism
%      25.70.-z, %Low and intermediate energy heavy-ion reactions
%      24.10.Lx, %Monte Carlo simulations (including hadron and parton cascades and string breaking models)
%      21.65.-f  %Nuclear matter
%      25.75.Ld %Collective flow

%      }
%\keywords{Relativistic retardation effect; symmetry energy; heavy-ion reactions, pion production, collective flow}
\maketitle

\section{Introduction}\label{introduction}
The momentum distribution of nucleons in a nuclear system, as a direct reflection of strong interactions, has always been a fascinating topic in nuclear physics~\cite{Czy60,Sar80,AHP88,Pand97,Egi,Frank08}. Qualitatively, some consensuses on the nucleon momentum distribution (NMD) have been reached, i.e., a large proportion of nucleons occupy the low-lying nuclear states with the momentum no more than Fermi momentum $k_{F}$, while the rest minority of nucleons form a high momentum tail (HMT) in the NMD due to short-range correlations (SRCs)~\cite{AHP88,Pand97}. Experimentally, the knock-out reactions~\cite{Acla99,Malk01,Tang03,Shne07,Fomin12} have also confirmed that nucleons in the HMT are short-range correlated. Moreover, the $np$ dominance of SRCs is further found through measurements of relative abundances of $np$, $nn$ and $pp$ pairs in nuclei from $^{12}$C to $^{208}$Pb in high energy electron scattering experiments at Jefferson Laboratory (JLab)~\cite{Subedi08,Ohen14,Duer19}. Theoretically, the dominance of $np$ over $pp$ pairs is attributed to the existence of tensor forces in the $np$ deuteron-like state~\cite{Sar05,Alvi08}, and this is also confirmed by experimental analysis findings~\cite{Pia06} and theoretical calculations using various Monte Carlo method~\cite{Schi07,Cruz19}. Quantitatively, the experimental results at JLab suggest that about 20\% nucleons are in the HMT in a nucleus from $^{12}$C~\cite{Subedi08} even to $^{208}$Pb~\cite{Ohen14,Ohen18}. Also, the systematic analyses of these results at JLab indicate that the fraction of nucleons in the HMT is about 25\% in the symmetric nuclear matter (SNM) at the saturation density $\rho_{0}$~\cite{Ohen14,Ohen15a,Ohen15b}, however, the theoretical calculations using various many-body theories become to deviate significantly from this fraction for the HMT in the SNM at $\rho_{0}$, see, e.g., Refs~\cite{Rios09,LiZH16} for more details.

The momentum correlation function (MCF) of nucleon-nucleon (NN) pairs is widely used to study particle emissions and collision dynamics as well as anomalous structures of the halo nuclei ~\cite{Lync83,Poch87,Boal90,Gong91,Bauer92,Ieki93,Hand95,Colo95,Hein99,Wied99,Marq01,Verd02,Chen03L,Ghet04,Chen03,Chen04,Ma06,Cao12}. Of particular interest, authors in Refs.~\cite{Chen03,Chen04} and~\cite{Ma06} employing the Boltzmann-Uehling-Uhlenbeck (BUU) model and Quantum-Molecular-Dynamics (QMD) model, respectively, studied effects of the symmetry energy $E_{sym}(\rho)$ on the MCF of NN pairs under various beam energies, they found that the $E_{sym}(\rho)$ affects the MCF significantly in light reaction systems; while with the reaction system becoming heavier and/or the beam energy increasing, effects of the $E_{sym}(\rho)$ on the MCF become negligible. Moreover, consistent with the study in Ref.~\cite{Bauer92}, they found that the MCF of NN pairs, especially the proton-proton ($pp$) pairs, is less influenced by the in-medium NN cross sections. Stimulated by these studies, we demonstrate in heavy reaction systems with high beam energies that the MCF of $pp$ pairs from preequilibrium emissions can be as an effective probe to the HMT of the NMD due to it responds sensitively to the presence as well as fraction of nucleons in the HMT of the NMD but is almost insensitive to other factors including the $E_{sym}(\rho)$ and the uncertainty of cutoff value of nucleon effective high momentum.
\section{The Model}\label{Model}
In this study, an isospin- and momentum-dependent BUU transport model \cite{Das03,IBUU} is used as the event generator. However, to obtain reliable phase-space information after the last strong interaction, i.e., freeze-out, we have improved our model including the consideration of the pion potential and the isospin-dependent $\Delta$ potential~\cite{Wei19a} as well as fitting the high-momentum behaviors of the nucleon optical potential extracted from nucleon-nucleus scattering experiments~\cite{Chen14b,LXH13} and distinguishing the density dependencies of the in-medium $nn$, $pp$ and $np$ interactions~\cite{Wei18a,Wei18b,Wei19a}. Specifically, the nuclear interaction~\cite{Wei18a,Wei18b,Wei19a} of this model is expressed as
\begin{eqnarray}
U(\rho,\delta ,\vec{p},\tau ) &=&A_{u}(x)\frac{\rho _{-\tau }}{\rho _{0}}%
+A_{l}(x)\frac{\rho _{\tau }}{\rho _{0}}+\frac{B}{2}{\big(}\frac{2\rho_{\tau} }{\rho _{0}}{\big)}^{\sigma }(1-x)  \notag \\
&+&\frac{2B}{%
\sigma +1}{\big(}\frac{\rho}{\rho _{0}}{\big)}^{\sigma }(1+x)\frac{\rho_{-\tau}}{\rho}{\big[}1+(\sigma-1)\frac{\rho_{\tau}}{\rho}{\big]}
\notag \\
&+&\frac{2C_{l }}{\rho _{0}}\int d^{3}p^{\prime }\frac{f_{\tau }(%
\vec{p}^{\prime })}{1+(\vec{p}-\vec{p}^{\prime })^{2}/\Lambda ^{2}}
\notag \\
&+&\frac{2C_{u }}{\rho _{0}}\int d^{3}p^{\prime }\frac{f_{-\tau }(%
\vec{p}^{\prime })}{1+(\vec{p}-\vec{p}^{\prime })^{2}/\Lambda ^{2}},
\label{MDIU}
\end{eqnarray}%
and the corresponding parameters embedded in above expression are in forms of
\begin{eqnarray}
A_{l}(x)&=&A_{l0} - \frac{2B}{\sigma+1}\big{[}\frac{(1-x)}{4}\sigma(\sigma+1)-\frac{1+x}{2}\big{]},  \\
A_{u}(x)&=&A_{u0} + \frac{2B}{\sigma+1}\big{[}\frac{(1-x)}{4}\sigma(\sigma+1)-\frac{1+x}{2}\big{]}.
\end{eqnarray}
Here, the $E_{sym}(\rho)$ parameter $x$ affects only the isovector properties of the asymmetric nuclear matter (ANM) at nonsaturation densities. Generally, without the consideration of correlations in a nuclear system, the kinetic part of symmetry energy $E^{kin}_{sym}(\rho)$ is calculated from the free Fermi gas model as $E^{kin}_{sym}(\rho)=8\pi p^{5}_{f}/{9m h^{3}\rho}{\approx}12.5(\rho/\rho_{0})^{2/3}$ with $p_{f}=\hbar (3\pi^{2}\rho/2)^{1/3}$, it is however, under the consideration of SRCs, this expression should be modified because the $E^{kin}_{sym}(\rho)$ is reduced significantly due to SRCs according to some solid evidences from microscopic many-body theories~\cite{Lov11,Vida11,Car12,Rios14} as well as experimental analysis findings~\cite{Ohen14,Ohen15a,LiBA15}. Actually, as indicated in Ref.~\cite{Rios14}, the reduction of $E^{kin}_{sym}(\rho)$ is the only effect that we are able to identify as correlation-driven, and thus can be utilized as the sole criterion to incorporate the tensor force effects into nuclear effective interactions in a phenomenological manner. To this end, we can readjust the parameters embedded in nuclear interactions to phenomenologically incorporate the tensor force effects into nuclear effective interactions under the consideration of SRCs. In the actual readjustment, considering that a large proportion of nucleons are uncorrelated and only minority of nucleons are correlated, therefore, it is suitable to assume that the $E^{kin}_{sym}(\rho)$ also holds for the 2/3 regularity with respect to densities. Moreover, according to a microscopic Brueckner-Hartree-Fock (BHF) calculation using the Av18 interactions plus the Urbana IX three-body force~\cite{Vida11,Car12,Car14}, the $E_{sym}(\rho)$ at $\rho_{0}$ is almost completely contributed from its potential part. It is therefore we use an expression $E^{kin}_{sym}(\rho)=12.5{\big[}(\rho/\rho_{0})^{2/3}-1{\big]}$ for the $E^{kin}_{sym}(\rho)$ to meet these demands under the consideration of SRCs similar to previous studies~\cite{Yong16}. Using empirical constraints on properties of nuclear matter at $\rho_{0}=0.16$ fm$^{-3}$, i.e., the isoscalar constraints on the SNM including the binding energy $E_{0}(\rho_{0})=-16$ MeV, the incompressibility $K_{0}=230$ MeV, the isoscalar effective mass $m^{\star}_{s}=0.7m$, the isoscalar potential at infinitely large nucleon momentum $U_{0}^{\infty}(\rho_{0})=75$ MeV, as well as the isovector constraints on the ANM including the $E_{sym}(\rho_{0})=32.5$ MeV and the symmetry potential at infinitely large nucleon momentum $U_{sym}^{\infty}(\rho_{0})=-30$ MeV, we have fitted the parameters embedded in nuclear interactions for the scenarios with and without SRCs, respectively. The values of these parameters are shown in Table~\ref{tab1}, they are marked as w/o HMT and with HMT, respectively. Moreover, for the convenience of describing the $E_{sym}(\rho)$ using the parameter $x$, we have also shown the corresponding slope value $L$ of $E_{sym}(\rho)$ at $\rho_{0}$ in the bottom of Table~\ref{tab1}. It needs to be emphasized that the uncertainty of $E^{kin}_{sym}(\rho)$ as well as the total $E_{sym}(\rho)$ at nonsaturation densities and thus our used expressions for them do not change the results obtained in this paper due to the MCF of $pp$ pairs from preequilibrium emission is almost robustly insensitive to the $E_{sym}(\rho)$ in the studied reaction system as shown in the following parts.
\begin{center}
\begin{table}[th]
\caption{{\protect\small The parameters used in the present study and the corresponding $L$ of $E_{sym}(\rho)$ at $\rho_{0}$=0.16 fm$^{-3}$. }} \label{tab1}
\begin{tabular}{|c| c c|}
\hline
\quad parameters & \quad w/o HMT  & \quad with HMT \\
\hline
$A_{l0}$~(MeV) \quad & $-96.963$ \quad & $-96.963$ \\
$A_{u0}$~(MeV) \quad & $-36.963$ \quad & $-36.963$ \\
$C_{l}$~(MeV) \quad & $-40.820$ \quad & $-24.719$ \\
$C_{u}$~(MeV) \quad & $-119.368$ \quad & $-135.469$ \\
$B$~(MeV) \quad & $141.963$ \quad & $141.963$ \\
$\sigma$ \quad & $1.2652$ \quad & $1.2652$ \\
$\Lambda/p_{f}$ \quad & $2.424$ \quad & $2.424$ \\
\hline
$L(x=-1)$~(MeV) \quad & $149.309$ \quad & $181.183$ \\
$L(x=0)$~(MeV) \quad & $88.654$ \quad & $120.528$ \\
$L(x=1)$~(MeV) \quad & $27.999$ \quad & $59.872$ \\
$L(x=2)$~(MeV) \quad & $-32.657$ \quad & $-0.783$ \\
\hline
\end{tabular}%
\end{table}
\end{center}
\begin{figure}[th]
\centerline{\includegraphics[width=0.95\columnwidth]{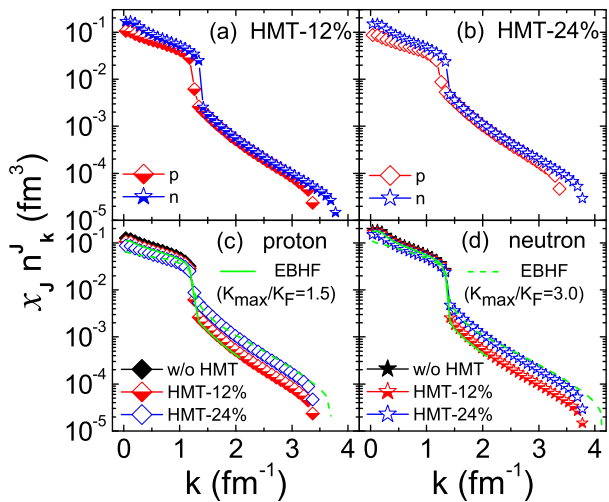}}
\caption{(Color online) Momentum distributions of a proton and neutron weighted by their respective fraction $x_{P}=Z/A$ and $x_{N}=(A-Z)/A$ in the initial $^{197}$Au nucleus without and with 12\% as well as 24\% high momentum nucleons in the HMT, and the corresponding distributions labeled as EBHF are obtained from Ref.~\cite{Yong19} using the local-density approximation. The normalization condition $\int_{0}^{\infty}4\pi k^{2}n^{J}_{k}{\rm d}{k}$=1 is used.} \label{mdis}
\end{figure}
\begin{figure*}[th]
\centerline{\includegraphics[width=0.8\textwidth]{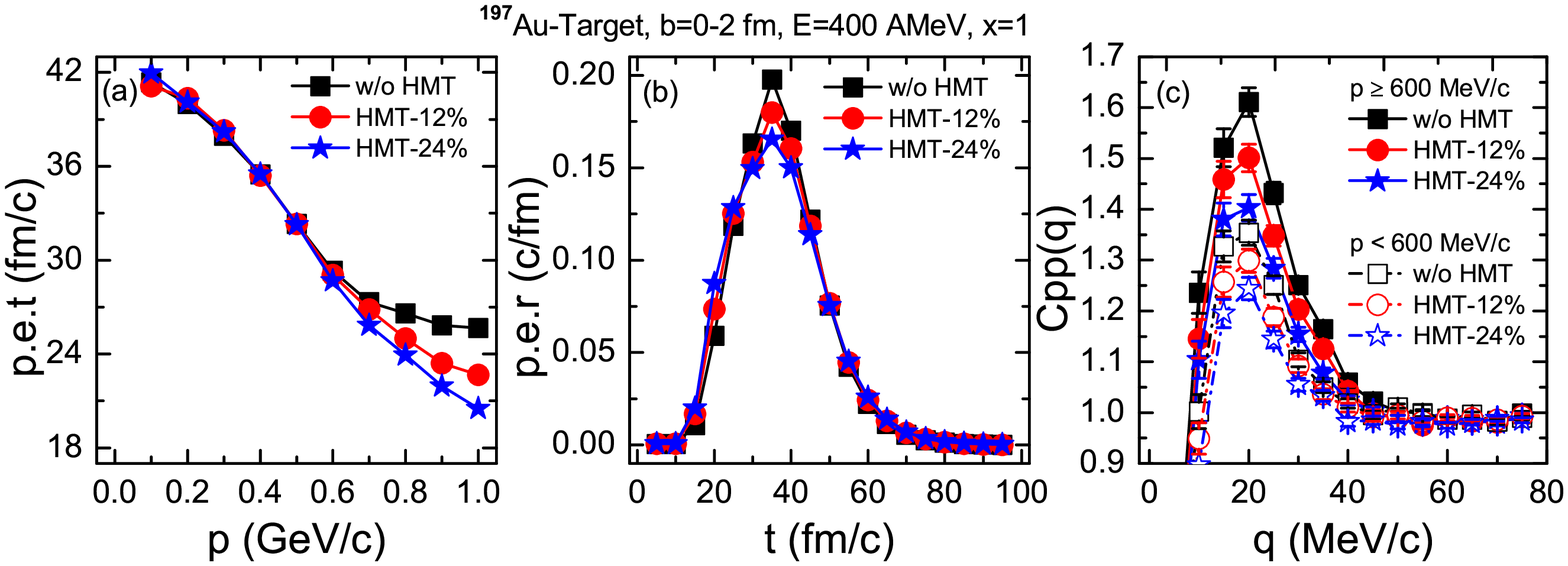}}
\caption{(Color online) Average proton emission time (p.e.t) and proton emission rate (p.e.r), and MCF of $pp$ pairs (CPP) without and with the SRCs, respectively.} \label{emission-crf}
\end{figure*}
\begin{figure*}[th]
\centerline{\includegraphics[width=0.8\textwidth]{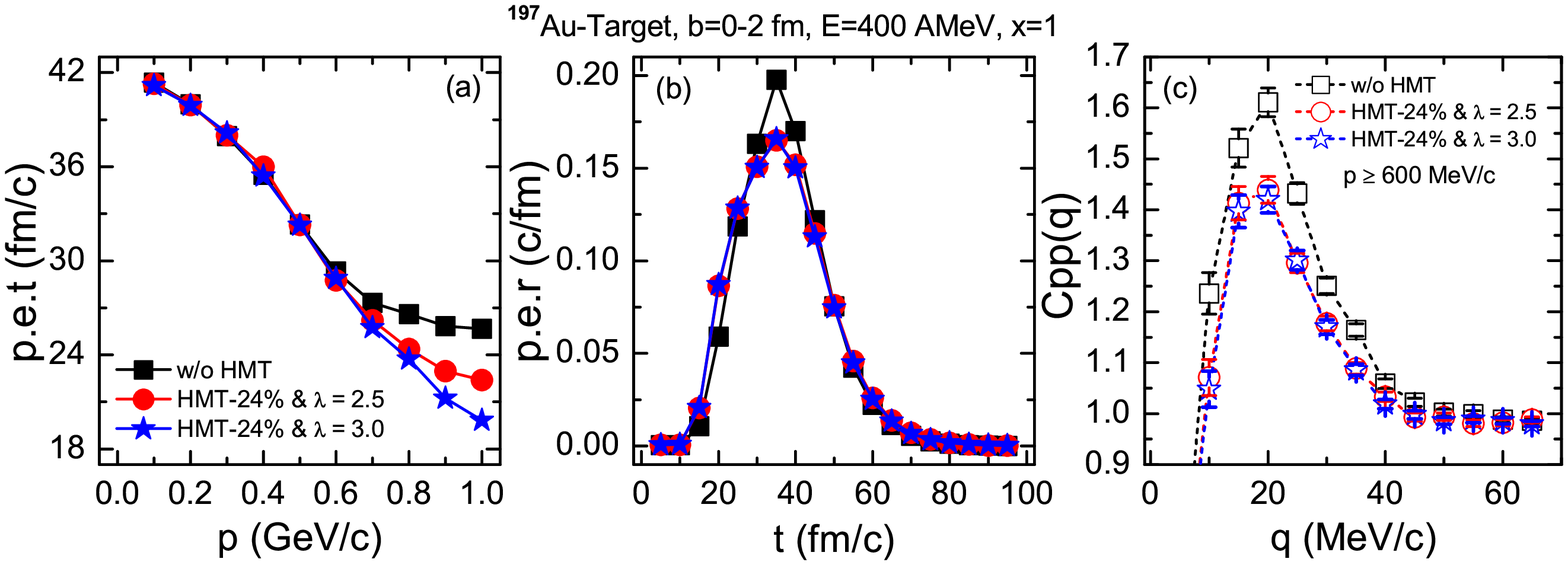}}
\caption{(Color online) The same as Fig.~\ref{emission-crf} but using two values for the $\lambda$ parameter, i.e., 2.50 and 3.00. } \label{lambda1}
\end{figure*}

For the specific form of the HMT, we use the $1/k^{4}$ distribution that has been confirmed in nuclei from $^{12}$C to $^{208}$Pb at JLab by the CLAS Collaboration~\cite{Ohen14,Ohen18}; while for the attainable maximum momentum of nucleons in the HMT, we use the form $k_{max}$=$\lambda k_{F}$, where the parameter $\lambda$=2.75$\pm$0.25 is the cutoff value of nucleon effective high momentum suggested by the experimental analysis findings~\cite{Ohen15a}. Except for specific illustrations, we will take the value 2.75 for the $\lambda$ parameter due to the MCF of $pp$ pairs is also robustly unchanged to the varies of the $\lambda$ parameter in the allowed range as shown in the following parts. In the actual initialization of $^{197}$Au nuclei, we also consider the isospin dependence for these high momentum nucleons. To this end, according to the recent experimental findings~\cite{Ohen18} about the isospin dependence of the high-momentum fraction for nucleons in neutron-rich nuclei, we adopt a relative fraction $N/Z$ for protons and approximate 1 for neutrons to initialize the high momentum nucleons. To compare effects of the fraction of nucleons in the HMT on the MCF, we take two values for the fraction of nucleons in the HMT to initialize the $^{197}$Au nuclei. First, we assume that 20\% of 118 neutrons in $^{197}$Au (i.e., 23.6 neutrons) have high initial momentum, then 20\%$\times$118/79$\approx$29.87\% of 79 protons (i.e., 23.6 protons) in $^{197}$Au also have high initial momentum, this yields a value 47.2/197$\approx$24\% for the fraction of nucleons in the HMT, and a value 5.99 for the SRC scaling factor, i.e., $a_{2}(A)$ which is independent of the momentum and is the probability of finding a high momentum $pn$ pair in nucleus $A$ relative to the deuterium~\cite{Egi,Frank08,Cruz19,Lynn19,Sar14,LiBA18}. As the second case, we assume that 10\% of 118 neutrons in $^{197}$Au (i.e., 11.8 neutrons) and the corresponding 10\%$\times$118/79$\approx$14.94\% of 79 protons (i.e., 11.8 protons) have high initial momentum, yielding a value 23.6/197$\approx$12\% for the fraction of nucleons in the HMT and a value 2.99 for the SRC scaling factor\footnote{The high momentum region from 300 to 600 MeV/$c$ for the $^{197}$Au nucleus is used to estimate the SRC scaling factor $a_{2}$ in this study, and the corresponding SRC probabilities used for the deuteron is about 4\%~\cite{Ohen15a}. For case of 24\% high momentum nucleons in the $^{197}$Au nucleus, the estimated value 5.99 of $a_{2}$ is within the predicted range 6.19$\pm$0.65 in Ref.~\cite{Wein11}, while for the case of 12\% high momentum nucelons in the $^{197}$Au nucleus, which is only a reference to compare effects of the fraction of nucleons in the HMT on the MCF, the estimated value 2.99 of $a_{2}$ is naturally far less than 6.19$\pm$0.65.}~\cite{Egi,Frank08,Cruz19,Lynn19,Sar14,LiBA18}. Obviously, the equal number of protons and neutrons with high momentum can ensure the $np$ dominance of SRCs in the HMT. Also, from the outputted momentum distribution of a proton and neutron weighted by their respective fraction $x_{P}=Z/A$ and $x_{N}=(A-Z)/A$ as shown in panels (a) and (b) of Fig.~\ref{mdis}, we can observe that the probabilities of finding the high momentum proton and neutron are approximate equal in the momentum range 250-600 MeV/$c$, i.e., $x_{P}\cdot n^{P}_{k}=x_{N}\cdot n^{N}_{k}$ in the range of $k$ from 1.26 to 3.05 fm$^{-1}$ due to in this momentum region the tensor interaction dominates in nuclei~\cite{Subedi08,Pia06}. In fact, this is exactly the first new property for high-momentum distribution of nucleons in neutron-rich nuclei indicated in Ref.~\cite{Sar14}. Certainly, one can also see that the neutron momentum distribution dominates the proton momentum distribution below the $k_{F}$, reflecting a fact that the probabilities of finding the neutron are larger than the proton because of more neutrons than protons are in this momentum region. Shown in panels (c) and (d) of Fig.~\ref{mdis} are the proton and neutron momentum distributions without and with 12\% as well as 24\% high momentum nucleons in the HMT, respectively. As comparisons, a parameterized isospin-dependent single NMD for the isospin ANM calculated from the extended BHF (EBHF) calculations~\cite{Yong19} is also used to initialize the $^{197}$Au nucleus using the local-density approximation~\cite{Gan94,Vive04}, it is seen that the corresponding momentum distributions under setting maximum momentum for high momentum nucleons in the range from 1.5$k_{F}$ to 3$k_{F}$ can approximately cover the range of our assumed NMD for both protons and neutrons. Moreover, we can also observe a low momentum depletion (LMD) below the $k_{F}$ and a corresponding HMT above the $k_{F}$ in the NMD, and as expected, this phenomenon is especially apparent for the scenario using the fraction 24\% for nucleons in the HMT. Therefore, these differences of NMD in $^{197}$Au+$^{197}$Au collisions are expected to enter through emission probabilities $g({\rm \bf p},x)$ of nucleons with momentum ${\rm \bf p}$ from the space-time point $x=({\rm \bf r},t)$ into the two-nucleon MCF evaluated by the standard Koonin-Pratt equation~\cite{Koon77,Kpf},
\begin{equation}\label{Kpf}
C({\rm \bf P},{\rm \bf q})=\frac{\int d^{4}x_{1}d^{4}x_{2}g({\rm \bf P}/2,x_{1})g({\rm \bf P}/2,x_{2})|\phi({\rm \bf q},{\rm \bf r})|^{2}}{\int d^{4}x_{1}g({\rm \bf P}/2,x_{1})\int d^{4}x_{2}g({\rm \bf P}/2,x_{2})},
\end{equation}
where ${\rm \bf P}(={\rm \bf p}_{1}+{\rm \bf p}_{2})$ and ${\rm \bf q}(=\frac{1}{2}({\rm \bf p}_{1}-{\rm \bf p}_{2}))$ are the total and relative momenta of nucleon pairs, respectively; and $\phi({\rm \bf q},{\rm \bf r})$ is the two-nucleon relative wave function where their relative position is ${\rm \bf r}=({\rm \bf r}_{2}-{\rm \bf r}_{1})-\frac{1}{2}({\rm \bf v}_{1}+{\rm \bf v}_{2})(t_{2}-t_{1})$.
\section{Results and Discussions}\label{Results and Discussions}

Statically understanding, the presence of HMT and thus SRCs in the initial colliding nuclei, naturally, will lead to a reduction of correlation emissions of the $pp$ pairs in the initial compression stage. As a result, compared to case without SRCs in the initial colliding nuclei, the MCF of $pp$ pairs evaluated from early emissions is expected to reduce in scenarios with SRCs. Indeed, as shown in panel (c) of Fig.~\ref{emission-crf}, this can be confirmed by comparing the corresponding MCF of $pp$ pairs with momentum per proton above about 600 MeV/$c$, which is calculated using the Pratt's correlation after burner (CRAB) code~\cite{Kpc} from the target in $^{197}$Au+$^{197}$Au collisions. Moreover, the larger fraction of high momentum nucleons in the initial colliding nuclei can cause the smaller probabilities of correlation emissions of the $pp$ pairs and thus the smaller values for the corresponding MCF of the $pp$ pairs in this period. However, the correlation between nucleons is dynamical during reactions, it is therefore we show in panels (a) and (b) of Fig.~\ref{emission-crf} a global scene of average emission time and emission rate for all protons from the target in $^{197}$Au+$^{197}$Au collisions. Obviously, the average emission time of protons with momentum $p$ greater than about 600 MeV/$c$ is rather sensitive to the presence as well as fraction of nucleons in the HMT as shown in panel (a) of Fig.~\ref{emission-crf}, i.e., compared to cases without and/or with fewer nucleons in the HMT, the case with more nucleons in the HMT gets these protons earlier emission due to the larger pressure generated in colliding region by more high momentum nucleons as well as their violent collisions. Also, more nucleons in the HMT lead to more protons emitting in this period, this can also be confirmed by comparing the corresponding average emission rate before the moment about 27 fm/$c$ as shown in panel (b) of Fig.~\ref{emission-crf}. Actually, at the early compression stage, as the projectile starts to approach and then gradually compresses the target, nucleons in target especially those in the HMT will have the larger probabilities to be accelerated into the region with momentum greater than about 600 MeV/$c$, this is the reason why we see in panel (b) of Fig.~\ref{emission-crf} the larger emission rate before the moment about 27 fm/$c$ in collisions with more nucleons in the HMT. Certainly, with the emissions of more high momentum nucleons in collisions with SRCs at the early compression stage, the unemitted nucleons in collisions with SRCs are naturally less than those in collisions without SRCs. As a result, for the later compression and expansion stages, the average emission rate in collisions with SRCs will be less than that in collisions without SRCs. Indeed, for the subsequent emissions of the protons, this can be confirmed by the smaller emission rate after the moment about 27 fm/$c$ with more nucleons in the HMT as shown in panel (b) of Fig.~\ref{emission-crf}. Nevertheless, it should be mentioned that the nucleon average emission time at the later compression and expansion stages is not obviously different in these two scenarios due to these unemitted nucleons are almost from the Fermi sea and thus approximately have identical average momentum. On the other hand, while we have readjusted the interaction used for the scenario with SRCs according to the sole criterion of correlation-driven as aforementioned, this is carried out phenomenologically at the level of the mean-field, and thus the off-shell effects of short-range correlated nucleons are not considered properly. Naturally, this interaction does not work well to directly derive the formation of correlation pairs during reactions. On the contrary, through comparing the MCF of $pp$ pairs with momentum per proton above about 600 MeV/$c$ with that below about 600 MeV/$c$, we can find that the initial correlation effects dominate the degree of correlation emissions of the $pp$ pairs. In other words, the observed decrease of MCF of the $pp$ pairs with momentum per proton above about 600 MeV/$c$ (i.e., preequilibrium emissions) in scenarios with SRCs is mainly a direct reflection of the depletion of nuclear Fermi sea of initial colliding nuclei. Naturally, as the second-order effects of initial correlation, the MCF of $pp$ pairs with momentum per proton below about 600 MeV/$c$ does not respond as sensitively to the presence as well as fraction of nucleons in the HMT as that above about 600 MeV/$c$ does as shown in panel (c) of Fig.~\ref{emission-crf}.

As far as the uncertainty of cutoff value $\lambda$ of nucleon effective high momentum, we can fix the fraction of nucleons in the HMT under the consideration of SRCs, and then check the sensitivity of the MCF to the $\lambda$ parameter in the allowed range. To this end, we set the fraction of nucleons in the HMT as 24\%, and take two values for the $\lambda$ parameter, i.e., 2.50 and 3.00 in calculations. Shown in panels (a) and (b) of Fig.~\ref{lambda1} are the corresponding average emission time and emission rate for all protons from the target in scenarios with and without SRCs. It is seen that setting $\lambda$ a larger value gets the nucleons in the HMT to have the higher initial momentum, and thus causes the earlier emission of the preequilibrium protons. Nevertheless, these more earlier emission does not affect the proton emission rate at both compression and expansion stages. Naturally, the uncertainty of cutoff value of nucleon effective high momentum does not affect the MCF of $pp$ pairs in probing the fraction of nucleons in the HMT as shown in panel (c) of Fig.~\ref{lambda1}. These results also imply that the emission rate plays an important role in determining the degree of correlation emissions. More importantly, it is seen that the nucleon emission rate before and even at the moment about 15 fm/$c$ is not more than 2.5\%, this level of stability of the ground state is good enough for the statistical results obtained here.

\begin{figure}[th]
\centerline{\includegraphics[width=0.95\columnwidth]{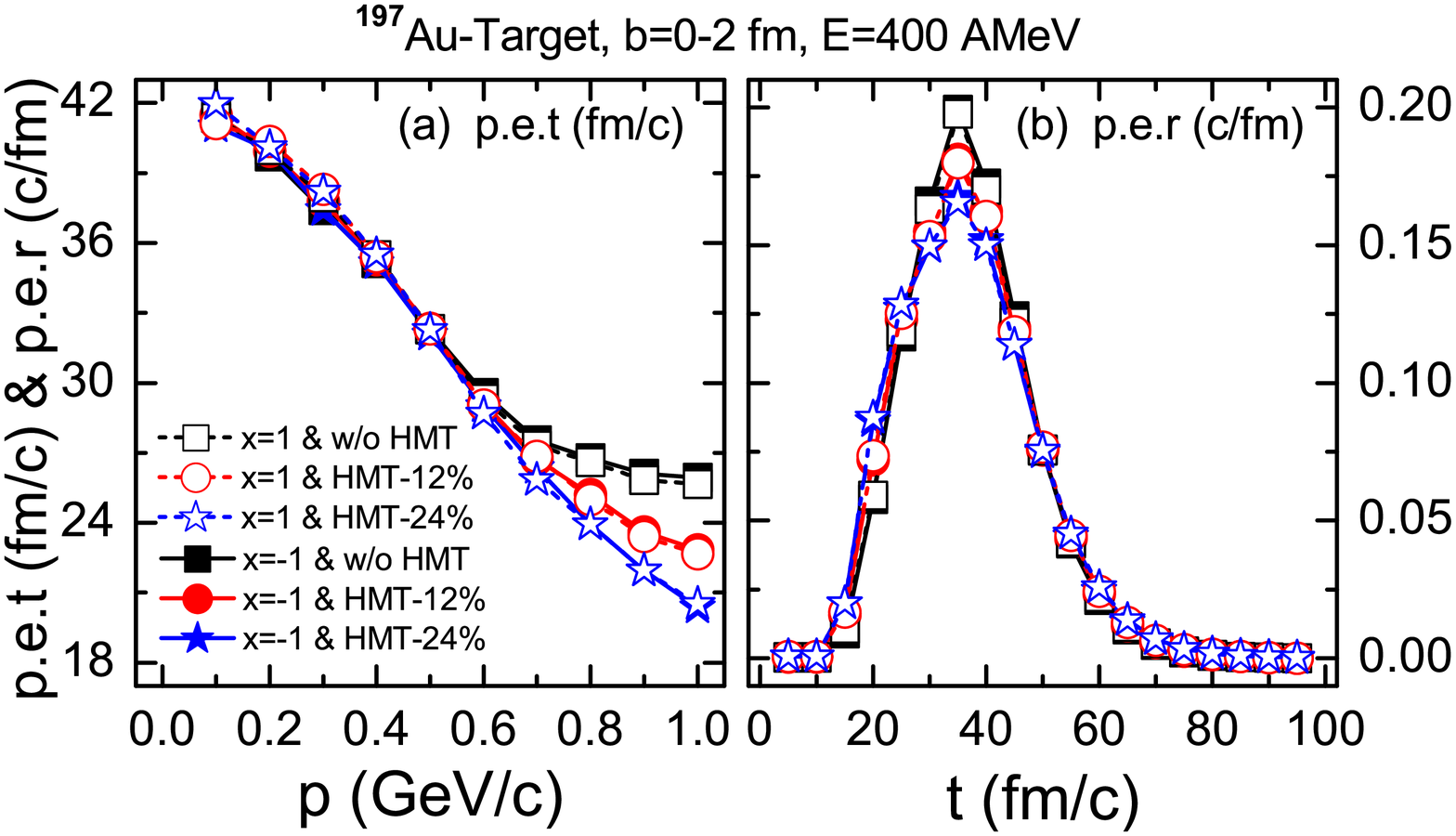}}
\caption{(Color online) Average proton emission time (p.e.t) and proton emission rate (p.e.r) without and with the SRCs, respectively.} \label{esym1}
\end{figure}
\begin{figure}[th]
\centerline{\includegraphics[width=0.95\columnwidth]{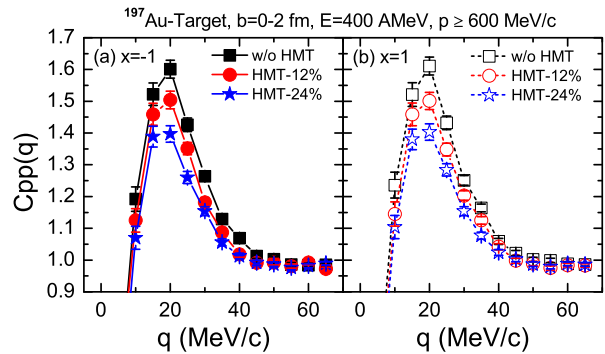}}
\caption{(Color online) MCF of the preequilibrium $pp$ pairs (CPP) with and without SRCs, respectively.}\label{esym2}
\end{figure}
\begin{figure}[th]
\centerline{\includegraphics[width=0.75\columnwidth]{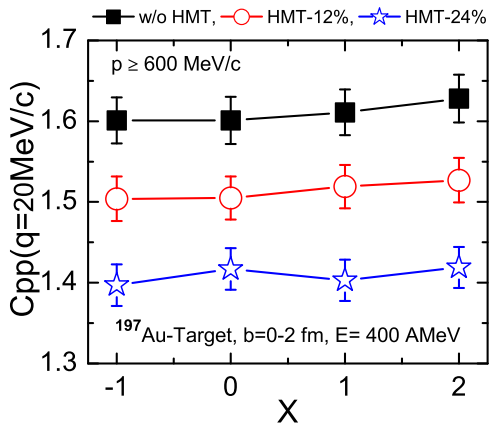}}
\caption{(Color online) The peak values of the MCF of $pp$ pairs from preequilibrium emission as a function of the $E_{sym}(\rho)$ with and without SRCs, respectively.}\label{esym3}
\end{figure}

Certainly, before regarding the MCF of $pp$ pairs from preequilibrium emissions as an effective probe to the fraction of nucleons in the HMT of the NMD, we still need to check the response of this observable to the $E_{sym}(\rho)$ because all the results are obtained from calculations using a specific $E_{sym}(\rho)$ with the parameter $x$=1. To this end, we show in Fig.~\ref{esym1} the average emission time and emission rate for all protons from the target in $^{197}$Au+$^{197}$Au collisions using a stiff $E_{sym}(\rho)$ with $x=-1$ and a soft one with $x=1$. It is seen that the $E_{sym}(\rho)$ does not affect both the average emission time and emission rate. Naturally, the corresponding MCF of $pp$ pairs evaluated from preequilibrium emissions is almost insensitive to the $E_{sym}(\rho)$ as shown in Fig.~\ref{esym2}. In fact, just as indicated in Refs.~\cite{Chen04,Ma06}, with the reaction system becoming heavier and/or the beam energy increasing, effects of the $E_{sym}(\rho)$ on two-nucleon MCF become negligible. Therefore, it is not surprising to see that the $E_{sym}(\rho)$ does not affect the MCF of $pp$ pairs in $^{197}$Au+$^{197}$Au collisions at 400 MeV/nucleon. Moreover, to clearly show sensitivities of the MCF of $pp$ pairs to the presence as well as fraction of nucleons in the HMT, we have also shown in Fig.~\ref{esym3} the peak values of the MCF of $pp$ pairs in scenarios with and without SRCs, in which the $E_{sym}(\rho)$ is set in a broader range with $x$ from $-1$ to $2$. It is obvious to see that the peak values of the MCF are rather sensitive to the presence as well as fraction of nucleons in the HMT, but are almost insensitive to the $E_{sym}(\rho)$. In terms of the sensitivity and clearness, it is therefore we suggest the MCF of $pp$ pairs from preequilibrium emissions as an effective probe to the fraction of nucleons in the HMT of the NMD.
\section{Summary}\label{Summary}
In conclusion, we have studied effects of the presence as well as fraction of nucleons of the HMT in the NMD on the MCF of $pp$ pairs within an improved transport model. It is shown that the presence as well as fraction of nucleons of the HMT in the initial NMD can lead to an appreciable reduction of correlation emissions of the preequilibrium $pp$ pairs. Moreover, the larger value of the fraction of nucleons in the HMT causes the smaller probabilities of correlation emissions of the preequilibrium $pp$ pairs. On the other hand, it is shown that the MCF of $pp$ pairs from preequilibrium emissions is almost robustly insensitive to the stiffness of $E_{sym}(\rho)$ as well as the uncertainty of cutoff value of nucleon effective high momentum. In terms of the sensitivity and clearness, we suggest the MCF of $pp$ pairs from preequilibrium emissions as an effective probe to the fraction of nucleons of the HMT in the NMD.

%\section*{\textbf{Acknowledgements}}
G. F. Wei would like to thank Profs. Bao-An Li, Gao-Chan Yong and Li Ou for helpful suggestion and discussion. This work is supported by the National Natural Science Foundation of China under grant Nos.11965008, 11405128, 11565010, U1731218, U1832129, and the PhD-funded project of Guizhou Normal university (Grant No.GZNUD[2018]11), and the innovation talent team (Grant No.(2015)4015)  and the high level  talents (Grant No.(2016)-4008)) of the Guizhou Provincial Science and Technology Department, and also in part by the Xi'an Science and Technology Planning project under grant No.CXY1531WL35.

\end{CJK*}

\end{document}